\documentclass[graybox,natbib,nosecnum]{svmult}
\bibpunct{(}{)}{;}{a}{}{,} % suppress commas between author-names and year

\pdfoutput=1   %forces use of pdflatex. Disable if you prefer to use .eps or .ps figures.
\usepackage{mathptmx}       % selects Times Roman as basic font
\usepackage{helvet}         % selects Helvetica as sans-serif font
\usepackage{courier}        % selects Courier as typewriter font
\usepackage{type1cm}        % activate if the above 3 fonts are
\usepackage{makeidx}         % allows index generation
\usepackage{graphicx}        % standard LaTeX graphics tool
\usepackage{multicol}        % used for the two-column index
\usepackage[bottom]{footmisc}% places footnotes at page bottom
\usepackage[normalem]{ulem}	% for strike-through of text with \sout{}  
\usepackage[colorlinks=true,linkcolor=blue,citecolor=blue,filecolor=blue,urlcolor=blue]{hyperref}

\usepackage[T1]{fontenc}

\usepackage{soul}   % for high-lighting of text
\newcommand{\hbindex}[1]{\hl{#1}\index{#1}}  %highlights index entries

\makeindex             % used for the subject index
\usepackage{amsmath,amssymb}
\usepackage{bm}
\usepackage{color}
\usepackage{xspace}
\usepackage{marginnote,setspace}
\usepackage{multirow}

\newcommand{\ccc}[1]{\marginnote{\setstretch{0.5}\textcolor{red}{\small \textit{#1}}}\textcolor{red}{$^\ast$}}

\newcommand{\Tb}[1]{\mbox{Table~\ref{tab:#1}}}
\newcommand{\fg}[1]{Fig.~\ref{fig:#1}}
\newcommand{\Fg}[1]{\mbox{Figure\ \ref{fig:#1}}}
\newcommand{\eq}[1]{Eq.~(\ref{eq:#1})}
\newcommand{\eqp}[1]{\mbox{(Eq.\ \ref{eq:#1})}}
\newcommand{\Eq}[1]{\mbox{Equation~(\ref{eq:#1})}}

\newcommand{\eqs}[2]{Eqs.\ (\ref{eq:#1}) and (\ref{eq:#2})}

\renewcommand{\ccc}[1]{}

\begin{document}

\title*{Pebble Accretion}
\author{Chris W. Ormel}
\institute{Chris W. Ormel \at Department of Astronomy, Tsinghua University, 100084 Beijing, China, \email{chrisormel@tsinghua.edu.cn}}
\maketitle

\abstract{Pebble accretion refers to the growth of planetary bodies through the accretion of pebble-sized particles. Pebbles are defined in terms of their aerodynamical size $\tau_s$, which describes the level of coupling to the disk gas. Observations confirms the presence of pebble-sized particles in both protoplanetary disks and the early solar system. Pebble accretion proceeds through the settling mechanism, where particles settle to the surface of the planet. This Chapter discusses the key aspects of the pebble accretion framework: the accretion regimes, the planet initiation mass, and the planet isolation masses. The accretion behavior of loosely coupled $\tau_s>1$ particles, referred to as ``large pebbles'', is also examined. The pebble accretion probability, $\epsilon$, is shown to be a useful parameter for evaluating the efficiency of the process, though this quantity is not necessarily high. Distinctions between pebble and planetesimal accretion are outlined. Pebble accretion, in particular, can be a highly effective mechanism in dense rings, as witnessed with ALMA.  
}

\section{Introduction: Pebbles}

Pebble accretion describes the growth of large bodies, ranging from planetesimals to planets, through the accretion of small, aerodynamically active particles \citep{OrmelKlahr2010}. In this process, the planetary body is massive enough to gravitationally attract solid particles, but not yet massive enough to rapidly accrete gas. Pebbles are particles of intermediate size: small enough for gas drag to play a decisive role \textit{during} the encounter but large enough to decouple from the gas. Pebble accretion provides a means to bridge the gap between ${\sim}$km-sized planetesimal bodies and the ${\sim}10\,m_\oplus$ mass at which runaway gas accretion commences. Pebble accretion should not be confused with processes that describe the formation of macroscopic bodies itself, which could also involve pebble-sized particles. In particular, the \hbindex{streaming instability} (or other drag-mediated instabilities) describes how planetesimal size bodies form by concentration and subsequent gravitational collapse. But this is not pebble accretion. 

Before further introducing pebble accretion we first must clarify what pebbles are in an astrophysical context.  I will define
\begin{svgraybox}
    Pebbles are aerodynamically active particles that can drift over significant distances within the lifetime of the protoplanetary disk
\end{svgraybox}
In the next, I review the observational evidence for the existence of pebble-sized particles in disks and  how pebbles are defined from an aerodynamical perspective. The remainder of this Chapter introduces the pebble accretion framework as it has developed over the past decade.

\subsection{Observational Evidence for Pebbles in Disks}
In protoplanetary disks, pebbles settle to the midplane regions due to the vertical component of the stellar gravity. The midplane regions of the disk would typically be inaccessible with facilities that operate at optical and IR wavelengths (including JWST). It is only at radio wavelength that the midplane regions are revealed. Here, ALMA has revolutionized the field by spatially resolving disks, showing a variety of substructure in the forms of spirals, arcs, and, especially, rings \citep{BaeEtal2023}. This substructure is best exhibited at continuum wavelengths, which traces the underlying solids population.  The prevalence of substructure offers us with the first line of evidence for the dominance of pebble-sized particles in disks. Unless substructure can directly be generated from infalling material \citep{KuznetsovaEtal2022,CalcinoEtal2024} only pebble-sized particles would have the aerodynamical properties to migrate over the distances required to display the substructure.

A straightforward example of this argument is that disks tend to be seen significantly smaller in continuum than in the gas \citep{SanchisEtal2021,LongEtal2022}. The explanation is that pebble-sized particles have vacated the outer disks, drifting inwards over distances ${\sim}10{-}100\,\mathrm{au}$. 
A more subtle example is that pebble-sized particles, would strongly react to any local pressure perturbation, i.e., pebbles pile-up in pressure maxima. Even when the magnitude of the pressure perturbation that lies behind the reversal is small, $|\delta P|/P \ll 1$, it would greatly amplify the imprint on the pebble disk. The at times spectacular substructure seen in ALMA continuum requires aerodynamically active particles.

The second line of evidence for the existence of pebbles, is the spectral dependence of the observed emission $I_\nu$ at sub(mm)/radio wavelengths. This dependence arises from the size dependence of the opacity $\kappa_\nu$---the absorption cross section per unit solid mass. In the Interstellar Medium (ISM), grains are thought to be small, reaching sizes up to ${\sim}0.1\,\mu\mathrm{m}$. At sub(mm) wavelengths the Rayleigh limit is applicable and the opacity decreases with wavelength, $\kappa_\nu \propto \nu^{\beta}$.  In the ISM $\beta\approx1.7$. On the other hand, particles of size much larger than the wavelength would \textit{not} exhibit a wavelength dependence ($\beta=0$). Hence, grain growth would manifest itself as a decrease in $\beta$. 

%Observationally, $\beta$ can be constrained from the wavelength dependence of the intensity $I_\nu$. 
Therefore, if the emission is optically thin and thermalized the spectral index $\alpha$ (as in $I_\nu\propto\nu^\alpha$) is 1-1 related to $\beta$ since $I_\nu = \tau_\nu B_\nu(T) = \kappa_\nu \Sigma B_\nu(T)$. If the temperature is known (and under the optically thin assumption) measuring $\alpha$ informs us on the sizes of the underlying particles.
%where $\Sigma$ and $T$ are the surface density and temperature of the emitting material, $\tau_\nu$ the opacity, and $B_\nu(T)$ the Planck spectrum. In the Rayleigh limit, $B_\nu \propto \nu^2$ and therefore $I_\nu \propto \nu^{2+\beta}$. In other words, measuring the frequency index of the intensity ($\alpha=2+\beta$) informs us about the size of the particles. 
Indeed, studies indicate $\beta$ values substantially smaller than the ISM value, providing evidence for grain growth to pebble sizes \citep{TestiEtal2014}. Although there are many complicating factors to this simple reasoning (optical depth and scattering effects, temperature dependence, dependence of material properties and internal structure) the consensus is that the change in $\alpha$ reflects grain growth \citep{SierraEtal2021,Guerra-AlvaradoEtal2024}. The inferred disk mass in pebbles for the typical T-Tauri disk are quite small, perhaps only several Earth masses \citep{AnsdellEtal2017}. Mass budgets could be much higher, however, if the emission is optically thick \citep{RicciEtal2012i} or when scattering effects matter \citep{ZhuEtal2021}. Mass budgets are also considerably higher in younger Class 0/I sources \citep{TychoniecEtal2020}.

Other complementary evidence for the presence of grain growth comes from millimeter polarization measurements in disks \citep{KataokaEtal2015,YangEtal2016}. The significant polarization fraction of the ALMA bands 6 and 7 ($\lambda\approx1\,\mathrm{mm}$) seen in several disks betray the presence of pebbles in the ${\approx}150\,\mu\mathrm{m}$--mm size range, dependent on their porosity \citep{TazakiEtal2019}.

In the Solar System evidence for the existence of pebbles comes from the bodies that have preserved their interior structure. Undifferentiated (chondritic) meteorites are often seen stuffed with sub-mm sized spherules known as chondrules---${\sim}300\,\mu\mathrm{m}$-sized particles that underwent flash-heating events \citep{MarrocchiEtal2024}. If chondrules (or their progenitor particles) predate the parent bodies, rather than being the collisional product of planetary bodies \citep{ConnollyJones2016}, and if chondritic meteorites represent a dominant population, this would strongly suggest that pebble-sized particles were indeed abundant in the early Solar System. Further support for the presence of pebble-sized particles comes from space missions targeting asteroids and comets. Data from multiple experiments conducted by the \textit{Rosetta} mission revealed that Comet 67P/Churyumov–Gerasimenko is composed of pebbles ranging 3 to 6 mm in radius \citep{BlumEtal2017}.

\subsection{Pebble definition}

According to the Udden-Wentworth scale pebbles are particles between 4 and 64 mm in diameter \citep{WilliamsEtal2006}. \citet{Wentworth1922} further writes:
\begin{svgraybox}
``Pebble is from the Anglo-Saxon papol, which meant something small and round, perhaps akin to the Latin \textit{papula, a pustule}.''
\end{svgraybox}
More recently, the term pebble has been introduced in the context of pebble accretion \citep{LambrechtsJohansen2012}. Pebbles in this context are neither defined in terms of size nor necessarily round. Instead, the implied meaning of pebble is aerodynamical: pebbles are particles that couple in a certain way to the gas. 

Particles moving through a gaseous medium experience gas drag. The amount of gas drag is customarily expressed in terms a drag coefficient $C_D$:
\begin{equation}
    \bm{F}_\mathrm{drag} = -\frac{1}{2} C_D A \rho_\mathrm{gas} v\bm{v}
    \label{eq:Fdrag}
\end{equation}
where $A$ is the cross-sectional area of the particle, $\bm{v}$ the velocity of the particle with respect to the gas, $v=|\bm{v}|$, and $\rho_\mathrm{gas}$ the gas density. The lower the $C_D$, the more economical your sports car or jet aircraft. 

At low densities, two points should be notes. First, the drag coefficient $C_D$ in \eq{Fdrag} is not constant, but instead varies with the particle Reynolds number, defined $\mathrm{Re}_p = 2sv/\nu_\mathrm{mol}$, where $\nu_\mathrm{mol}$ is the molecular viscosity and $s$ the particle radius (see, e.g., \citealt{Weidenschilling1997} for expressions valid for spherical particles). For $\mathrm{Re}_p<1$ drag is linear in velocity. Second, when the mean free path of gas molecules ($l_\mathrm{mfp}$) exceeds the particle radius, the drag force follows the specular reflection limit:
\begin{equation}
    \bm{F}_\mathrm{drag} = - \frac{4\pi}{3}s^2 \rho_\mathrm{gas} v_\mathrm{th} \bm{v} \qquad (s<\frac{9}{4}l_\mathrm{mfp})
    \label{eq:Fdrag-molecular}
\end{equation}
where $v_\mathrm{th}=\sqrt{8kT/\pi m_\mathrm{gas}}$ is the gas mean thermal motion.  In this limit drag is also linear in velocity. This means that a particle's \hbindex{stopping time} $t_\mathrm{stop}\equiv mv/|F_D|$ becomes independent of velocity: 
\begin{equation}
    t_\mathrm{stop} = \left\{ \begin{array}{l@{\qquad}l@{\qquad}l}
            \displaystyle \frac{\rho_\bullet s}{\rho_\mathrm{gas} v_\mathrm{th}}  & s<\frac{9}{4}l_\mathrm{mfp} & \textrm{Epstein} \\[1.4em]
            \displaystyle \frac{\rho_\bullet s}{\rho_\mathrm{gas} v_\mathrm{th}} \left( \frac{4s}{9l_\mathrm{mfp}} \right)  & \frac{9}{4}l_\mathrm{mfp}<s< s[\mathrm{Re}_p=1] & \textrm{Stokes} \\
    \end{array}
    \right.
    \label{eq:tstop}
\end{equation}
The independence of stopping time on velocity is the reason $t_\mathrm{stop}$ is used to quantify the aerodynamical interaction of a particle. Often the stopping time is nondimensionalized through to the orbital frequency
\begin{equation}
    \tau_s = t_\mathrm{stop}\Omega_K
\end{equation}
a quantity that is referred to as the particles' \hbindex{aerodynamical size}\footnote{In the literature this quantity is often referred to as "Stokes number".}. Thus, $\tau_s<1$ indicates particles adjust to the gas flow within one orbital period, while $\tau_s>1$ particles do not. The latter can be described within the framework of Keplerian orbits, with gas drag acting as a perturbative force. For $\tau_s\ll1$ particles the Kepler representation becomes meaningless, as these particles closely follow the motion of the gas. This makes $\tau_s=1$ particles the fastest drifters (see below). Pebbles are therefore defined in terms of the aerodynamical stopping time $\tau_s$. Depending on the context, pebbles can be classified as particles with $10^{-2} \le \tau_s \le 10^2$ or in some cases $10^{-3} \le \tau_s \le 10^3$. Particles with $\tau_s>1$ are referred to as \hbindex{large pebbles}. Also, any particle in the absence of gas ($\tau_s\rightarrow\infty$) is not a pebble from an aerodynamical perspective.

\subsection{Drift motions}
Protoplanetary disks are supported against the stellar gravity by rotation and pressure $P$. If rotation would be the only support mechanism (i.e., no pressure gradient) the orbital velocity of the gas would equal the Keplerian velocity $v_{\phi,\mathrm{gas}} = v_K = \sqrt{Gm_\star/r}$. In general, the radial pressure gradient (usually negative) provides some hydrostatic support, such that that the gas moves at a velocity $v_\mathrm{\phi,gas}\simeq (1-\eta)v_K$ where $\eta$ is defined 
\begin{equation}
    \eta \equiv - \frac{1}{2\Omega_K^2 r \rho_\mathrm{gas}} \frac{\partial P}{\partial r}
    \sim \frac{P}{v_K^2 \rho_\mathrm{gas}}
    \sim h_\mathrm{gas}^2.
    \label{eq:eta}
\end{equation}
Here, $\partial P/\partial r$ is assumed to be a power-law of radius. The gas disk aspect ratio is defined $h_\mathrm{gas}=c_s/v_K=H/r$, where $c_s$ is the isothermal sound speed and $H$ the gas pressure scaleheight. Hence, $\eta\ll1$, indicating that the gas rotates at near-Keplerian speeds, with a slight deviation from the circular Keplerian velocity by $\eta v_K \sim 10{-}100\, \mathrm{m\,s^{-1}}$. 

Particles do not feel the gas pressure ($\rho_\bullet \gg \rho_\mathrm{gas}$) but they do interact with the gas aerodynamically through drag forces. Gas friction extracts angular momentum from the particles, which therefore tend to drift in radially. Solving for the steady solutions to the Euler equations for gas and particles simultaneously, we find that the drift velocities, in a frame co-moving at the Keplerian orbital velocity are \citep{NakagawaEtal1986}:
\begin{equation}
    \label{eq:vdrift}
    v_r     = -\frac{2\tau_s}{\tau_s^2 +(1+Z_\mathrm{mid})^2} \eta v_K; \qquad
    v_\phi  = -\frac{(1+Z_\mathrm{mid})}{\tau_s^2 +(1+Z_\mathrm{mid})^2} \eta v_K
\end{equation}
where $Z_\mathrm{mid}$ is the local (midplane) solids-to-gas ratio. Since we will typically assume in this chapter that a planet moves on a circular orbit these are the velocities at which the planet faces the pebbles (see \fg{geom}). For small $\tau_s$ and $Z_\mathrm{mid}\approx0$ we have that $v_\phi\simeq -\eta v_K$, which expresses that pebbles move along with the gas, and $v_r\simeq -2\eta v_K \tau_s$. The radial drift peaks at $\tau_s=1$ at which point the drift timescale $t_\mathrm{drift}=r/|v_r| \sim 1/\eta\Omega_K$, is just several hundreds of orbital periods. The drift velocities decrease for either $\tau_s\gg1$ or $Z_\mathrm{mid}\gg1$. In the case where $\tau_s<1$ but $Z_\mathrm{mid}\gg1$ this happens because the tightly-coupled particles drag the gas along to Keplerian rotation.

\begin{figure}[t]
    \includegraphics[width=\textwidth]{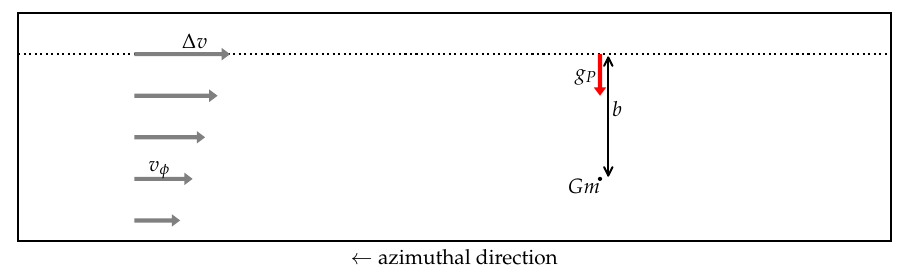}
    \caption{\label{fig:geom}Geometry of the encounter between planet and pebble. The planet moves on a circular orbit in the positive azimuthal direction (leftwards), facing pebbles at a relative velocity $v_\phi$---the headwind term. For large impact parameters $b$ the approach velocity $\Delta v$ is determined by the Keplerian shear. In reality, the approach velocity $\Delta v$ could also have contributions from $v_r$, planet eccentricity, and turbulence. At closest approach, the pebble experiences a gravitational acceleration $g_P = Gm/b^2$.}
\end{figure}

\section{The physics of pebble accretion}
I will give two definitions of pebble accretion:
\begin{svgraybox}
    \begin{enumerate}
        \item Pebble accretion is an accretion mechanism where a planet accretes pebble---particles that can drift over large distances (as defined above).
        \item Like above, but where in addition the pebbles are \textit{captured} in the planet's gravitational well and settle to the surface of the planet. Such particles move on \textit{settling trajectories}. Settling trajectories require that the pebble-planet interaction time $t_\mathrm{enc}$ exceeds the pebble's stopping time $t_\mathrm{stop}$.
    \end{enumerate}
\end{svgraybox}
The second definition is usually what is understood with ``pebble accretion''. I will describe these settling trajectories first, before outlining the region in the planet mass--pebble size parameter space where we can find these.

\subsection{Settling trajectories}
\begin{figure}[t]
    \includegraphics[width=\textwidth]{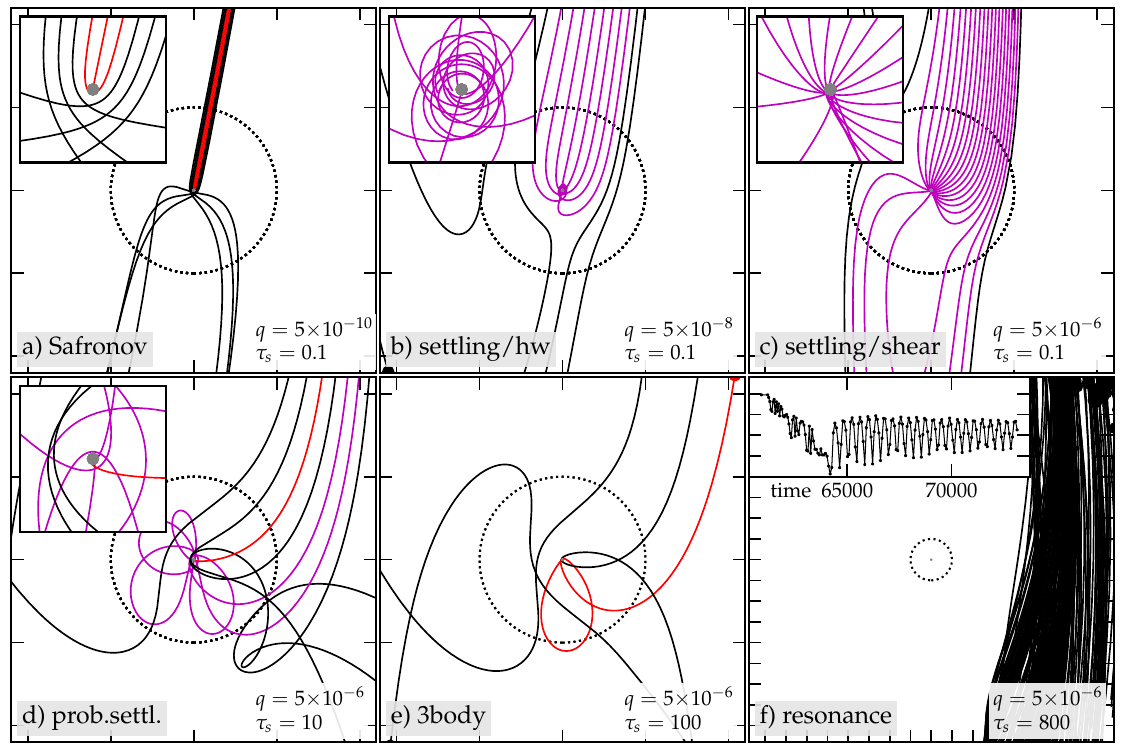}
    \caption{\label{fig:mosaic}Trajectories of pebbles with aerodynamical size $\tau_s$ in a frame co-moving with the planet of mass $q=m/m_\star$ on a circular orbit. Trajectories in red indicate ballistic accretion: particles hit the surface of the planet (here taken to be $R/R_h=5\times10^{-3}$). Trajectories in purple indicate settling: these particles would be captured in the Hill radius and be accreted even when $R{\rightarrow}0$. The gas flow pattern is assumed to be unaffected by the planet. Dotted circle denotes the Hill radius and tick marks are also separated by $1\,R_\mathrm{H}$. Particles are launched from a ring outside the planet's orbit at fixed azimuthal intervals. Therefore, each trajectory carries the same amount of the pebble flux. c) some particles accrete from the back. e) Two trajectories are shown; the non-accreting particle re-enters the Hill radius three times. f) One particle trajectory is shown, which get trapped in resonance. The inset shows the distance to successive conjunctions as function of time.}
\end{figure}

The region about the planet dominated by its gravity, rather than that of the star, is given by its \hbindex{Hill radius}
\begin{equation}
    R_H = r \left( \frac{m}{3m_\star} \right)^{1/3}
    \label{eq:Rhill}
\end{equation}
where $m$ is the mass of the planet and $r$ the distance to the star of mass $m_\star$.  Pebble accretion can be defined as the \textit{capture} of particles within the Hill sphere (the second definition above). Such trajectories are referred to as \hbindex{settling trajectories}. These contrast accretion by \hbindex{ballistic interactions}, which rely on particles hitting a surface. Ballistic trajectories can graze the planet's surface, but such close encounters are absent if the accretion proceeds by settling.  The difference between settling and ballistic accretion paths is illustrated in \fg{mosaic}  with purple and red trajectories, respectively, for a variety of planet masses and particle stopping times. Would the surface disappear (e.g., the gravitating body shrinks to a point mass), ballistic accretion vanishes; in \fg{mosaic}, trajectories in red would then exit the Hill radius. Particles accreted through the settling mechanism, on the other hand, would be unaffected. This definition is not only academic. In the Hill units adopted in \fg{mosaic}, the planets surface, $R/R_H$, shrinks with stellar orbital radius $r$ and most ballistic accretion vanishes (red trajectories turn black). Pebble accretion is therefore comparatively more relevant in the outer disk.

Gas drag balancing gravity will result in a terminal velocity---the settling velocity---equal to $v_\mathrm{settl}=g_P t_\mathrm{stop}$, where $g_P$ is the gravitational acceleration of the planet.  For pebbles to settle towards the planet, the settling timescale
\begin{equation}
    t_\mathrm{settl} = \frac{b}{v_\mathrm{settl}} = \frac{b^3}{Gm t_\mathrm{stop}}
    \label{eq:tsettl}
\end{equation}
with $b$ the impact parameter (see \fg{geom}), must be shorter than the duration of the encounter, $t_\mathrm{enc}$. The latter could be estimated as ${\sim}b/\Delta v$ where $\Delta v$ is the velocity at which a pebble approaches the planet. Equating $t_\mathrm{settl}$ and $t_\mathrm{enc}$ we readily find the accretion rate%can solve for $b^2 \Delta v$, resulting in an accretion rate of
\begin{equation}
    \label{eq:mdot-3d}
    \dot{m}_\mathrm{3D} \sim \rho b^2 \Delta v \sim Gm t_\mathrm{stop} \rho.
\end{equation}
This is the 3D limit, because we assumed that particles are homogeneously distributed over the entire cross section ${\sim}b^2$ at density $\rho$. Remarkably, $\dot{m}_\mathrm{3d}$ is independent of the relative velocity $\Delta v$. If the particle layer is sufficiently settled with respect to the impact parameter $b$, the accretion becomes 2D. In that case we solve for $b$ as function of $\Delta v$ first to obtain:
\begin{equation}
    \label{eq:mdot-hw}
    b\sim \sqrt{\frac{Gmt_\mathrm{stop}}{\Delta v}}
    \quad \textrm{and} \quad
    \dot{m}_\mathrm{2d} 
    \sim b\Delta v \Sigma 
   %\sim \sqrt{\frac{Gmt_\mathrm{stop}}{\Delta v}} \Delta v \Sigma
    \sim \sqrt{ Gm t_\mathrm{stop} \Delta v} \Sigma.
\end{equation}
Note that we have not specified the source of the approach velocity $\Delta v$; it can be due to the drift of the particles or the eccentricity of the planet (or something else).  The limit where $\Delta v$ is determined by the pebble drift, $\Delta v \sim |v_\phi| \sim \eta v_K$, is referred to as the \hbindex{headwind limit} or the Bondi limit.  If, on the other hand, the approach velocity is given by the Keplerian shear, $\Delta v \sim b\Omega_K$ and $t_\mathrm{enc}\sim\Omega_K^{-1}$, we instead obtain
\begin{equation}
    \label{eq:mdot-shear}
    \dot{m}_\mathrm{2d-shear} \sim \left( \frac{Gm t_\mathrm{stop}}{\Omega_K} \right)^{2/3} \Omega_K \Sigma \sim R_h^2 \Omega_K \tau_s^{2/3} \Sigma
\end{equation}
which is known as the \hbindex{shear limit} or Hill limit. In \fg{mosaic} panels (b) and (c) illustrate pebble accretion in the headwind and shear limits, respectively. It is in the 2D/shear limit where accretion rates peak.  For $\tau_s \sim 0.1$--$1$ almost all particles entering the Hill radius are accreted! 

There is an important caveat to the expressions derived above. In these, we have assumed particles approach their terminal settling velocities ($v_\mathrm{settl}=g_P t_\mathrm{stop}$). In reality, this takes a time $t_\mathrm{stop}$, meaning the interaction duration must exceed the stopping time ($t_\mathrm{enc}>t_\mathrm{stop}$)---a condition referred to as the \hbindex{settling condition}. Using $b$ from \eq{mdot-hw} and $t_\mathrm{enc}=b/\Delta v$, the settling condition translates to
\begin{equation}
    t_\mathrm{stop} \lesssim \frac{Gm}{(\Delta v)^3}
    \quad \textrm{or} \quad
    \tau_s \lesssim \frac{m}{m_\star} \left( \frac{\Delta v}{v_K} \right)^{-3}
    \quad \textrm{or} \quad
    m\gtrsim m_\ast \equiv \left( \frac{\Delta v}{v_K} \right)^3 \tau_s m_\star.
    \label{eq:mast}
\end{equation}
Planets with mass ${\ll}m_\ast$ cannot accrete pebbles through the settling mechanism (\fg{mosaic}a); the pebbles approach the planet too quickly. Pebble accretion requires slow encounters. Additionally, for pebbles $\tau_s \ge 1$ the encounter time is ${\sim}\Omega_K^{-1} < t_\mathrm{stop}$ and the settling condition to generally fail (see below). This occurs because it takes too long for such bodies to be significantly influenced by gas drag.

\begin{figure}[t]
    \includegraphics[width=\textwidth]{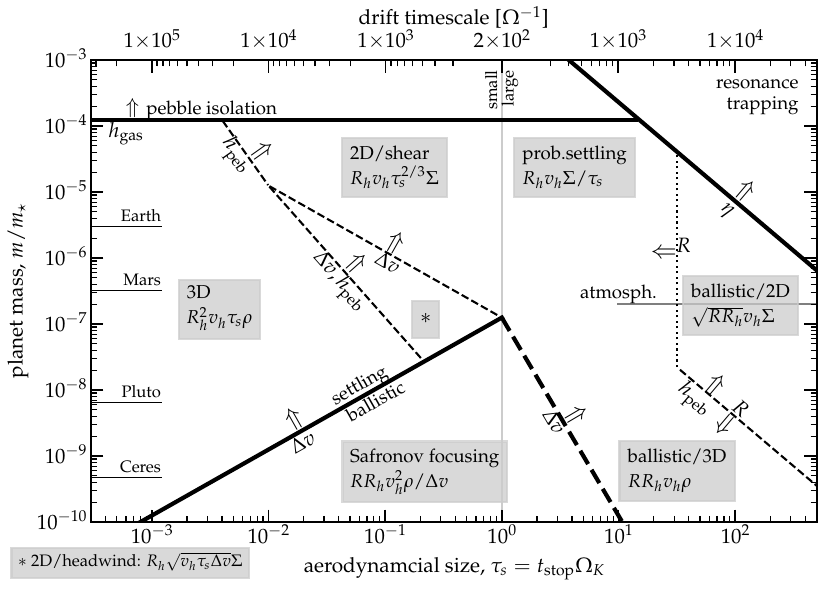}
    \caption{\label{fig:regimes}Schematic summarizing accretion mechanisms and their accretion rates $\dot{m}$.  The leading terms in the expression for $\dot{m}$ for each of the regimes are given.  Boundaries between accretion regimes are indicative and will in reality shift when accounting for numerical prefactors. It is further indicated how the regime boundaries shift upon change of the parameters $\Delta v$, $h_\mathrm{peb}$, $R$, and $\eta$. Here, $R$ is the physical radius of a body, $R_h$ its Hill radius, $v_h=R_h\Omega_K$, $\Delta v$ the approach velocity between planet and pebble, and $\rho$ the mass density of the pebbles. The figure has been drawn for $\eta=5\times10^{-3}$, $R/R_h=10^{-3}$, $h_\mathrm{peb}$ following \eq{hpeb} with $\delta_z=10^{-4}$ and a disk aspect ratio of $h_\mathrm{gas}=0.05$.  
    Different line styles indicate the nature of the transitions: thick solid lines represent abrupt transition (e.g., ballistic to settling), dashed lines denote gradual transitions (e.g., the 3D-to-2D transitions), and dotted lines signify ``fluid'' transitions, meaning that two distinct accretion mechanisms operate on both sides of the line.     Thick lines indicate abrupt changes in the accretion rate (e.g., onset of pebble accretion, pebble isolation, and resonances). The thick dashed line indicates ballistic trajectories are continuous across it, whereas settling trajectories are not.
    The dotted line at $\tau_s\approx30$ indicates that settling and ballistic accretion mechanisms operate on each side of the line.}
\end{figure}

\subsection{Pebble accretion parameter space}
\Fg{regimes} presents an overview of pairwise accretion mechanisms for both pebble and ballistic interactions.  
In \fg{regimes} and similar figures I assume that a (proto-)planet of mass $m$ (y-axis) on a circular orbit interacts with pebbles of aerodynamical size $\tau_s$ (x-axis), drifting inward according to the free drift expressions \eqp{vdrift}. The boundaries between the regimes are approximate as transition effects and numerical prefactors are not accounted for; they could easily shift by an order of magnitude in a certain direction.

In the small pebble limit ($\tau_s<1$) pebble accretion is limited by the pebble isolation mass from above and by the settling condition \eqp{mast} from below. Here settling encounters abruptly give way to the usual 2-body ballistic accretion (\fg{mosaic}a):
\begin{equation}
    \dot{m} = \pi R^2 (1 + \Theta) \Delta v \rho 
    = \pi R^2 \left( 1 + \frac{2Gm}{R(\Delta v)^2} \right) \Delta v \rho
    \label{eq:mdot-bal3d}
\end{equation}
where $\Theta$ is the gravitational focusing factor. The corresponding expression in \fg{regimes} only keeps the part including $\Theta$ and is written in Hill units where $Gm=3R_Hv_H^2$ and $v_H = R_H \Omega_K$. This ``Safronov focusing'' is central to many studies about planetesimal accretion where $\Delta v$ arises from the eccentricity of the planetesimal bodies. Safronov focusing for pebbles (i.e, the bottom triangle in \fg{regimes}) has $\Delta v$ given by \eq{vdrift}. 

We refer to pebbles with $\tau_s>1$ as \hbindex{large pebbles}.  By definition, large pebbles have stopping times longer than an orbital time and do not generally obey the settling condition.  Nevertheless, for not too large $\tau_s$ accretion can still proceed and in fact be dominated by settling (\fg{mosaic}d). While most particles get ejected from the Hill radius \textit{some} fail to escape the Hill radius to settle towards the planet. The efficiency of this \textit{probabilistic settling} decreases with $\tau_s^{-1}$. This implies that ballistic accretion, which operates in the 2D equivalent of \eq{mdot-bal3d} will gradually overtake settling at $\tau_s \sim \sqrt{R_h/R}$.  The transition is gradual; both accretion types co-exists for $\tau_s>1$ particles (as illustrated in \fg{mosaic}d). 

Accretion can be boosted greatly by the emergence of planet atmospheres. This occurs when the surface escape velocity exceeds the thermal velocity of the gas or when the Bondi radius
\begin{equation}
    R_B = \frac{Gm}{kT/m_\mathrm{gas}}
\end{equation}
exceeds $R$. The pre-planetary atmosphere is connected to the disk but is denser than the disk gas. This opens up the possibility for larger particles to be captured. I will nonetheless refer to this type of accretion as ballistic, because a planet's atmosphere is after all connected to the planet. The interaction can then be described with an atmosphere-enhanced capture radius $R_\mathrm{cap}(m,\tau_s,\dots)$ replacing the physical radius \citep{InabaIkoma2003}. 

Larger pebbles are even less influenced by gas and drift slowly towards the planet. As they approach, they experience long-range gravitational interactions with the planet they approach. The inward drift can then be offset by resonant repulsion. These pebbles are small enough to drift over significant distances but sufficiently large to be captured in mean motion resonances (\fg{mosaic}f).

\section{The pebble accretion framework}
I will discuss here the key elements that make up the pebble accretion framework. These include: accretion rates in the small and large pebble regimes, the pebble isolation mass and the pebble accretion initiation masses.

\subsection{Pebble accretion rates ($\tau_s<1$)}

\begin{table}[t]
    \renewcommand{\arraystretch}{1.2}
    \centering
    \caption{\label{tab:PA-rates}Asymptotic expressions for pebble accretion rates $\dot{m}$ and efficiencies $\epsilon$ of small pebbles ($\tau_s<1$) in the 3D and 2D limits.}
    \begin{tabular}{l@{\qquad}l@{\qquad}l@{\qquad}l@{\qquad}}
    \hline\noalign{\smallskip}
             &   $\displaystyle \frac{\dot{m}}{\Sigma r^2 \Omega_K}$     & $\displaystyle \frac{\dot{m}}{\rho r^3 \Omega_K}$ & $\epsilon$                \\[1em]
        \hline \noalign{\smallskip}
        3D limit ($\dot{m}_\mathrm{3D}$) & $4.9 \tau_s q /h_\mathrm{peb}$ & $12\tau_s q$ & $0.39q/\eta h_\mathrm{peb}$  \\
    \multicolumn{4}{l}{2D limit ($\dot{m}_\mathrm{2D}$)} \\
    -- headwind         &   $4.0 (\eta \tau_s q)^{1/2}$     &   & $0.32 (q/\eta\tau_s)^{1/2}$       \\                                          % &   $\eta < (q/\tau_s)^{1/3}$ \\
    -- eccentricity     &   $3.5 (e \tau_s q)^{1/2}$        &   & $0.28(eq/\eta^2\tau_s)^{1/2}$     \\            %  $e<(q/\tau_s)^{1/3}$ \\ %   & \\
    -- shear            &   $2.9 (\tau_s q)^{2/3 }$         &   & $0.23(q^2/\eta^3\tau_s)^{1/3}$    \\ %  $\eta +e \lesssim (q \tau_s)^{1/3} $  \\  %& \\
\noalign{\smallskip}\hline\noalign{\smallskip}
\end{tabular}
\flushleft
\small
\textbf{Notes.} Normalized accretion rates and efficiencies are given in terms of: $e$ (\textit{planet} eccentricity), $h_\mathrm{peb}$ (pebble aspect ratio \eqp{hpeb}), $\eta$ \eqp{eta}, $q = m/m_\star$,  and $\tau_s = t_\mathrm{stop}\Omega_K$ (aerodynamical size).  Expressions follow \citet{LiuOrmel2018} and \citet{OrmelLiu2018} and are valid only for $\tau_s \le 1$ and masses above $m_\ast$. In the 2D limit expressions are listed according to their approach velocity; the velocity limit that evaluates to the highest rate has precedence. In the 3D limit pebbles are assumed to be normally distributed according to scaleheight $H_\mathrm{peb}=h_\mathrm{peb}r$. In general, the 2D and 3D expressions should be combined according to \eq{m-combined}. 
\end{table}

Pebble accretion rates for $\tau_s<1$ particles are compiled in \Tb{PA-rates}. These are given in non-dimensionless form with the accretion rates expressed in units of either $\Sigma r^2 \Omega_K$ or $\rho r^3 \Omega_K$. The prefactors have been obtained from three-body simulations carried out by \citet{LiuOrmel2018} and \citet{OrmelLiu2018} and are valid in the stated asymptotic limits. In the 2D limit, accretion rates depend on the relative velocity $\Delta v$, which can be due to the headwind of the pebbles, the shear in the disk, or the eccentricity of the planet (whichever is highest).
In the 3D limit, accretion rates are independent of $\Delta v$. Here, rates are given in terms of density or, equivalently, $\Sigma$ and the pebble scaleheight (or their aspect ratio, $h_\mathrm{peb}=H_\mathrm{peb}/r$).   If the planet is on an inclined orbit, such that the magnitude of the vertical oscillations exceed the pebble scaleheight, the effective aspect ratio of the pebbles should be cast in terms of the planet inclination \citep{OrmelLiu2018}. 

In the 2D limit the pebble accretion rate that materializes is the maximum of the headwind, eccentricity, and shear limits. For the headwind and eccentricity limits, rates scale with the square-root of velocity ($\eta$ or $e$), reflecting a higher supply rate of pebbles. But this comes at the expense of a higher initiation mass (\eq{mast}). Accretion rates are the highest in the 2D, shear limit. Converting the rates into a growth timescale ($m/\dot{m}$), we find:
\begin{equation}
    \label{eq:2D-shear}
    t_\mathrm{grw}^\mathrm{2d/shear} = 2\times10^4\,\mathrm{yr}\ \left(\frac{m}{m_\oplus}\right)^{1/3} \left(\frac{m_\star}{m_\odot}\right)^{-5/6} \left(\frac{\tau_s}{0.01}\right)^{-2/3} \left( \frac{\Sigma r^2/m_\star}{10^{-5}} \right)^{-1} \left( \frac{r}{\mathrm{5\,au}}\right)^{3/2}.
\end{equation}
It is clear that if \eq{2D-shear} materializes, protoplanets can easily grow to the runaway gas accretion threshold mass. The ``problem'' with these high accretion rates is that the local reservoir of pebbles will be quickly emptied and needs to be resupplied from the outer disk. For these reasons, it is often more useful to express the accretion rates in terms of an accretion probability $\epsilon$ (see below).

\Tb{PA-rates} shows that none of the pebble accretion rates are superlinear with mass $q$. Two equal-mass embryos will grow similarly in terms of mass, a situation resembling oligarchic growth rather than runaway growth. However, runaway growth effects (i.e., one body outcompetes another one in terms of growth) can yet occur, e.g., when the two bodies are on opposite sides of the $R_\mathrm{init}$ threshold or in rings, where competitors can be scattered out of the pebble feeding zone \citep{LauEtal2022,JiangOrmel2023}.

\runinhead{Modulation factors}In reality transitions are smooth. \citet{OrmelLiu2018} found that the combined rate can be well represented from the respective 2D and 3D expressions
\begin{equation}
    \label{eq:m-combined}
    \dot{m} = \left(\dot{m}_\mathrm{2D}^{-2} +\dot{m}_\mathrm{3D}^{-2} \right)^{-0.5}.
\end{equation}
In addition, the initiation of pebble accretion around the mass $m_\ast$ has been fitted by a modulation factor:
\begin{equation}
    f_\mathrm{set} 
    = \exp \left[ -0.5 \left( \frac{\Delta v}{v_\ast} \right)^2 \right]
    = \exp \left[ -0.5 \left( \frac{m_\ast}{m} \right)^{2/3} \right]
    \label{eq:fset}
\end{equation}
where $v_\ast = (q/\tau_s)^{1/3} v_K$ and $m_\ast$ was defined in \eq{mast} \citep{OrmelKlahr2010,LiuOrmel2018}. 2D rates should be multiplied by $f_\mathrm{set}$ and 3D rates by $f_\mathrm{set}^2$.

A further modulation occurs when pebbles exhibit a size distribution, such that the net accretion rate will be averaged over a particle size distribution. In particular, a planet accreting pebbles of different stopping time may have its mass simultaneously below (for large $\tau_s)$ or above (for small $\tau_s$) the initiation mass $m_\ast$ \citep{LyraEtal2023}. Accounting for a size distribution will hence ameliorate the transition onset of pebble accretion.  Turbulence, in which $\Delta v$ follows a distribution, will have a similar effect \citep{OrmelLiu2018}.

\subsection{Pebble initiation mass}
It was discussed before that pebble accretion kicks in suddenly at the mass $m_\ast$ \eqp{mast}.   When comparing the accretion rates across the pebble accretion initiation threshold line (see \fg{regimes}), we obtain 
\begin{equation}
    \frac{\dot{m}_\mathrm{PA,3d}}{\dot{m}_\mathrm{Safr}} 
    \sim \frac{\Delta v}{R \Omega_K} \tau_s
   %\sim \frac{R_h}{R}\frac{\Delta v}{v_H} \tau_s 
    \sim \frac{\tau_s^{2/3}}{\alpha_R}
    \label{eq:boost}
\end{equation}
where, for simplicity, I took the 3D limit and used $\alpha_R = R/R_\mathrm{Hill}$ to relate radius to mass and substituted \eq{mast}. \Eq{boost} illustrates that pebble accretion (settling mechanisms) significantly boosts accretion rates by factors of 10 to 100, particularly in the outer disk where $\alpha_R$ is small. Of course, the transition is not discontinuous. By fitting the intersection of ballistic and settling encounter regimes, \citet{VisserOrmel2016} found the crossover point to lie at 
\begin{equation}
    \label{eq:Rinit}
    R_\mathrm{init} \approx 450\,\mathrm{km}\ \frac{\Delta v}{50\,\mathrm{m\,s^{-1}}} \left( \frac{\rho_\bullet}{\mathrm{1\,g\,cm^{-3}}} \right)^{-0.36} \left( \frac{r}{\mathrm{au}} \right)^{0.42} \left( \frac{m_\star}{m_\odot} \right)^{-0.14} \tau_s^{0.28}
\end{equation}
where $\rho_\bullet$ is the internal density of the planetesimal. Hence, at $m=m_\mathrm{init}$ (the mass corresponding to $R_\mathrm{init}$) accretion rates will steeply rise, until pebble accretion is fully activated at $m\sim m_\ast$ \eqp{mast}. 

Planetesimals must exceed $R_\mathrm{init}$ in order to enter the pebble accretion regime. Bodies that satisfy the constraint could rapidly grow to much larger sizes, rendering the planetesimal distribution bimodal. Conversely, if no single planetesimal exceeds the initiation threshold, planetesimals should grow by other, slower processes---either by ballistic accretion of pebbles or traditional planetesimal-driven accretion. This constraint is particularly severe for the outer disk, because $R_\mathrm{init}$ may easily be as large as Pluto and alternative growth mechanism (i.e., planetesimal accretion) are also slow. Often, planet population synthesis models bypass the initiation threshold condition, by starting with a sufficiently high seed mass which already accretes pebbles efficiently in the 2D limit \citep[e.g.][]{JohansenEtal2019,SchneiderBitsch2021}. However, the true question lies in understanding how these massive bodies originally formed.

The situation changes when $\Delta v$ is small. In \eq{Rinit} we took $\Delta v$ consistent with a smooth disk. However, in case of a local pressure maximum ($\eta\approx0$) or a particle pileup (or traffic jam; $Z_\mathrm{mid}\gg1$) $\Delta v$ can be much lower \eqp{vdrift}. Under these conditions, the challenges posed by the initiation threshold are largely mitigated, allowing growth to proceed more rapidly (see below).

\subsection{Pebble isolation mass}
The inward drift of pebbles exterior to a planet's orbit is driven by the negative pressure gradient in the disk. When the pressure gradient reverses, the pebble flux terminates \citep{LambrechtsEtal2014}. Pressure reversals can be due to disk-related processes, but the planet can also induce them. As it opens a gap in the disk, a (positive) pressure gradient (negative $\eta$) develops exterior to the planet, such that pebbles' drift velocity would become positive. Hence, pebbles pile up at the location where the contribution of the planet-induced (positive) pressure gradient and the disk-induced (negative) pressure gradient cancel. 

The typical scale for gap opening is set when the Hill radius of the planet exceeds the disk scaleheight, which translates into $m\sim h_\mathrm{gas} ^3 m_\star$. Hence, we write
\begin{equation}
    m_\mathrm{iso} = 42\,m_\oplus \left(\frac{h}{0.05}\right)^3 \left( \frac{m_\star}{m_\odot} \right) f_\mathrm{fit} f_\mathrm{diff}.
\end{equation}
Here $f_\mathrm{fit}$ is a numerically-obtained modulation function that accounts for effects of disk viscosity and the background pressure gradient \citep{BitschEtal2018}
\begin{equation}
    f_\mathrm{fit} = 0.6 \left( 0.66 +0.34 \left( \frac{-3}{\log_{10} \alpha_\nu} \right)^4 \right) \left( 0.58 - 0.17 \left( \frac{\partial \log P}{\partial \log r}\right)_\mathrm{disk} \right).
    \label{eq:ffit}
\end{equation}
When disks are more viscous or the background pressure gradient is steeper, a gap is harder to open and the pebble isolation mass is larger. The term $f_\mathrm{diff}$ accounts for the diffusive nature of the dust particles' motions in turbulent flows. At this stage a planet does not fully open a gap, allowing small particles to filter through along with the gas. According to \citet{AtaieeEtal2018}, this increases the pebble isolation mass by an amount 
\begin{equation}
    f_\mathrm{diff} = 1 +0.2 \left( \frac{\sqrt{\alpha_T}}{h_\mathrm{gas}} \sqrt{4 +1/\tau_s^2} \right)^{0.7}.
\end{equation}

\subsection{Large Pebble accretion ($\tau_s>1$)}
It is commonly believed that growing particles will hit a growth or drift barrier, preventing them from exceeding $\tau_s=1$ \citep[e.g.][]{BirnstielEtal2012}. Still $\tau_s>1$ pebbles can occur in a variety of situations. They could result from collisions between planetesimal bodies. Due to the weaker gravitational bonding of smaller bodies, a collisional cascade could be triggered, grinding a significant fraction of the mass down to particle sizes small enough for gas drag to become effective. \citet{KobayashiEtal2011} estimate this ``fragment size'' at $\tau_s\sim20$ dependent on planet mass, location, and material properties.  In addition, small pebbles can become aerodynamically large when the gas depletes, i.e., during disk dispersal or when outgassing occurs in debris disks. It is therefore meaningful to investigate the accretion properties of large pebbles. 

Similar to small pebbles, we define large pebbles as freely drifting particles (\eq{vdrift}), neglecting perturbations from bodies other than the planet. 
Recently, \citet{HuangOrmel2023} conducted three-body integrations (star, planet, and pebble) to investigate the large pebble accretion regime, employing a global reference frame.  These simulations directly determine whether the pebble is accreted by the planet or continues to drift inwards, providing the accretion probability ($\epsilon$). The advantage of employing a global setup over a local one is that it prevents double (or multiple) counting of accreting pebbles.  For example, when $v_r\rightarrow0$ the same pebble could repeatedly enter the planet Hill radius (as was seen in \fg{mosaic}e). However, in the local setup, unless care is taken, there is no way to distinguish a first from a repeating encounter.
Additionally, in the global frame the resonant forcing of the pebbles by the planet can be followed.

\begin{figure}[t]
    \centering
    \includegraphics[width=0.8\textwidth]{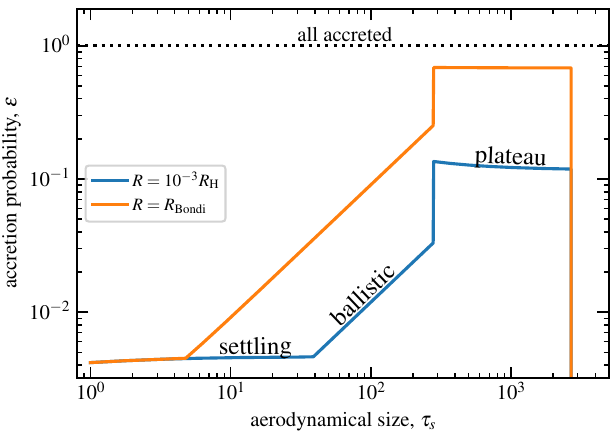}
    \caption{\label{fig:bigpebble}Probability that a drifting pebble gets accreted for large pebbles ($\tau_s>1$) with $q=10^{-6}$, $h_\mathrm{gas}=0.05$ and $\eta=5\times10^{-3}$.  Curves follow expressions by \citet{HuangOrmel2023}. The orange curve indicate the enhanced accretion probability due to the atmospheres, which may extend up to the Bondi limit (the exact enhancement depends on the atmosphere model). Beyond $\tau_s\approx3,000$ pebbles are captured in resonances.}
\end{figure}

\Fg{bigpebble} presents a summary of the findings by \citet{HuangOrmel2023}. The probability of getting trapped in the Hill radius by the settling mechanism is approximately ${\propto}\tau_s^{-1}$ reflecting the amount of energy dissipated within the Hill radius. At small $\tau_s$, the accretion \textit{probability} is therefore constant, because the reduction in the accretion rate is compensated by a slower drift. On the other hand, ballistic accretion rates stay constant with $\tau_s$ and in fact may increase greatly due to atmospheres (see below). \citet{HuangOrmel2023} found that the standard Safronov accretion rate
\begin{equation}
    \label{eq:mdot-bal}
    \dot{m}^\mathrm{2d} =  2R \Delta v \sqrt{1 + \frac{v_\mathrm{esc}^2}{\Delta v^2}} \Sigma
\end{equation}
fit their simulated data well.  Then, with increasing $\tau_s$ drift speeds become so slow that pebbles are guaranteed to enter the Hill radius. If not accreted at first instance, a pebble may reenter the Hill radius multiple times unless it either gets accreted or it is scattered to the inner disk. This results in a fixed accretion probability $\epsilon$, referred to as the ``plateau''.

Finally, for even larger $\tau_s$ particles end up in a first-order mean motion resonance (at period ratios $j{+}1{:}j$ with respect to the planet) and will no longer approach the planet. Here, the inward radial drift is compensated by the secular repulsion of the planet \citep{HasegawaNakazawa1990,MutoInutsuka2009}. The threshold stopping time $\tau_\mathrm{res}$ above which pebbles are captured in resonance depends on whether gas damping is ``strong'' or ``weak''. The weak limit, where the pebble will develop an eccentricity between successive conjunctions, usually applies for larger planet masses (as in \fg{bigpebble}). Importantly, the presence of a resonance barrier does not preclude accretion. As already noted by \citet{WeidenschillingDavis1985} particles trapped in these high-$j$ resonances 
would collide and fragment to smaller sizes. Approaching pebbles of aerodynamical size above $\tau_\mathrm{res}$ in fact are likely to be accreted in the plateau regime, in particular when an atmosphere is present.

\runinhead{Atmospheres} For planets with $R_B>R$, it has long been recognized that their atmospheres' can trap planetesimal bodies. \citet{InabaIkoma2003} showed that bodies impacting the planet at distance $R_c$, where the gas density $\rho(R_c)$ has been elevated due to the atmospheres, will lose enough energy to become trapped in the Hill sphere, when
\begin{equation}
    \rho(R_c) R_H \gtrsim \frac{6 + e_H^2}{9} s \rho_\bullet 
    \label{eq:R-capture}
\end{equation}
where $e_H$ is the Hill velocity (See \citealt{OkamuraKobayashi2021} for up-to-date discussion). If this criterion is satisfied we can simply replace $R_c$ for the planet's physical radius ($R$) in all preceeding expressions. Compared to planetesimals, large pebbles have lower $s$ and $e_H$. Therefore, \eq{R-capture} suggests that ballistic accretion rates for large pebbles would be greatly elevated, with $R_c$ increasing up to the Bondi radius (see \fg{bigpebble}). Accretion of large pebbles can be extremely efficient.

\subsection{Miscellaneous points}
There are other important facets to pebble accretion, which I will briefly touch on for completeness

\runinhead{Flow isolation}
The presence of a planet will alter the flow pattern of the gas in its vicinity, with the horseshoe motion a well-known manifestation. The deviation from the unperturbed flow pattern becomes particular pronounced at the scale of the Bondi radius $R_b$, i.e., when atmospheres appear. 3D hydrodynamical simulations typically show gas inflowing from the polar regions and an outflow of gas in the midplane \citep{OrmelEtal2015,CimermanEtal2017,KuwaharaKurokawa2024}. Small pebbles ($\tau_s\ll1$) that reside in the midplane could therefore be aerodynamically deflected away from the planet if $R_b$ exceeds the pebble accretion capture radius is. This flow isolation therefore occurs in the limit of small $\tau_s$ and large planets, although results are rather dependent on the gas drag law (Epstein or Stokes), gas thermodynamics (cooling), and the value of the radial pressure gradient (headwind or shear) \citet{KuwaharaKurokawa2020,KuwaharaKurokawa2020i}. \citet{KuwaharaKurokawa2020i} find that pebbles of $\tau_s \lesssim 10^{-3}$ generally avoid accretion for planets more massive than $R_b/H>0.03$.

\runinhead{Spin}
Particles that impact another body transfer angular momentum, contributing to its spin. While the spin of giant planets is believed to originate from gas accreted through a circumplanetary disk, the origin of the spin of the terrestrial bodies in the solar system is unclear. If the planetary bodies accreted mostly planetesimals, the imparted angular momentum would be insufficient \citep{DonesTremaine1993}. In contrast, accretion of pebbles transfers more substantial amounts of angular momentum \citep{VisserEtal2020}. However, once an atmospheres forms, its rotation is expected to become clearly prograde, which would drag the pebbles along \citep{TakaokaEtal2023}.

\runinhead{Pebble Torque}
Planets migrate in disks due to an axisymmetric distribution of density.  Even though solids constitute only $\sim$1\% of the mass, they can nevertheless induce an asymmetry that makes the net torque from solids comparable to the better-known gas-induced torques \citep{Benitez-LlambayPessah2018}. In particular, pebbles that are accreted from the front of the planet create a deficit in solids behind the planet's orbit  (see \fg{mosaic}b). This results in a positive and significant torque acting on the planet \citep{ChrenkoEtal2024}.

%This only covers the basics of pebble accretion, focusing mostly on the mechanical properties of pebble accretion. A lot has been neglected. 
\runinhead{Envelope enrichment and compositional implications}
A misconception of the core accretion model is that all accreted solids assemble in the central ``core'' of the planet. In reality, solids---especially slowly settling pebbles---can undergo thermal ablation in the hot envelopes surrounding accreting protoplanets \citep{VenturiniEtal2015,BrouwersEtal2018}. If the sublimated vapor remains in the envelope, it can expedite the onset of runaway gas accretion by increasing the mean molecular weight and decreases the adiabatic index of the gas \citep{BrouwersOrmel2020}. 

However, if ices sublimate already in the planet's upper atmosphere the vapor could be transported back to the disk along with the gas \citep{WangEtal2023,SteinmeyerEtal2023}. When this ``volatile recycling`` occurs, a compositional dichotomy between planet and disk may emerge: the planet retains the refractory elements from pebbles, while the disk gas becomes enriched in volatiles \citep{JiangEtal2023}.

\section{Pebble accretion as a planet formation mechanism}
Having established the basis of pebble accretion, we now turn to its ability to grow planets. A useful metric is the pebble accretion efficiency, which directly informs us on the required mass budget of pebbles. I will then discuss the different characteristics of pebble accretion in comparison to planetesimal accretion.

\subsection{The Pebble accretion efficiency}
\label{sec:efficiency}

Pebble accretion is not necessarily efficient in smooth disks, where pebbles can rapidly drift inward, missing the chance to encounter the planet. The \textit{probability} of a pebble to be accreted
\begin{equation}
    \epsilon \simeq \frac{\dot{m}}{\dot{M}_\mathrm{flux}}
    = \frac{\dot{m}}{\Sigma r^2 \Omega_K} \times \frac{v_K}{2\pi v_r}
    \label{eq:epsilon}
\end{equation}
can be low even though the accretion rate ($\dot{m}$) is high. For instance, if $\dot{M}_\mathrm{flux}=10^2\,\mathrm{m_\oplus\,Myr^{-1}}$, and $\epsilon=0.01$ a planet with mass 1 Earth mass doubles its mass in 1 Myr. But during this time $99\,m_\oplus$ in pebbles have drifted past the planet and are no longer available to its growth. This loss would be ameliorated if multiple planetary bodies reside in the disk to filter the pebble flux \citep{GuillotEtal2014}. Nevertheless, $\epsilon$ measures the viability of pebble accretion to produce planets. For planets to grow effectively, the preferred situation is that a few bodies consume \textit{most} of the pebbles, not that all pebbles get distributed over thousands of embryos.

\begin{figure}[t]
    \includegraphics[width=\textwidth]{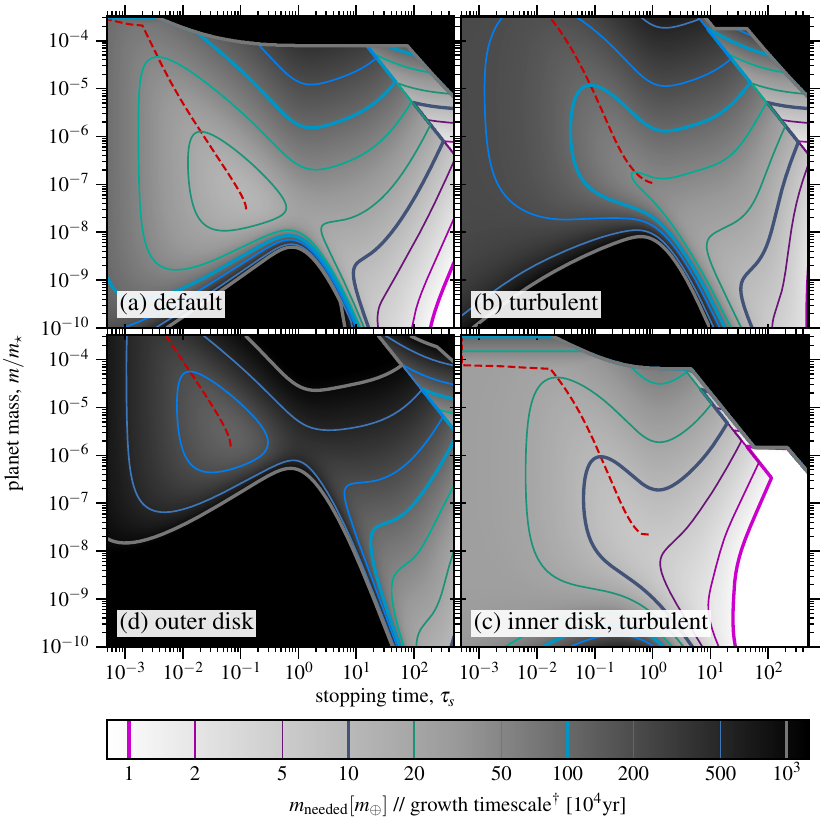}
    \caption{\label{fig:epsmosaic}Required pebble mass $M_\mathrm{needed}=m/\epsilon$ and growth timescale $t_\mathrm{growth}$ for pebble-accreting protoplanets in smooth disks. Contour levels assume a solar-mass star and a pebble flux of $100\,m_\oplus\,\mathrm{Myr}^{-1}$, in which case $M_\mathrm{needed}$ expressed in Earth units is equal to $t_\mathrm{growth}$ expressed in units of $10^4\,\mathrm{yr}$. Contour labels are proportional to stellar mass and (for the growth timescale) inversely proportional to the pebble mass flux $\dot{M}_\mathrm{flux}$. In (a) the efficiency has been calculated for $\delta_z=10^{-4}$, $\eta=5\times10^{-3}$, $h_\mathrm{gas}=0.05$ and $\alpha_R=10^{-3}$, representing conditions at $5\,\mathrm{au}$.  (b) Turbulent disk at 5 au. (c) Inner disk, also turbulent. (d) The outer disk. Here, \textit{turbulent} implies that $\delta_z=10^{-2}$; \textit{inner disk} that $\alpha_R=2\times10^{-2}$, $h_\mathrm{gas}=0.03$, and $\eta=10^{-3}$; and \textit{outer disk} that $h_\mathrm{gas}=0.1$, $\eta=2\times10^{-2}$, and $\alpha_R=10^{-4}$. The dashed red line indicates the location where $\epsilon$ peaks with respect to $\tau_s$.}
\end{figure}

In the last step of \eq{epsilon}, I have inserted $\dot{M}_\mathrm{flux} = 2\pi r v_r \Sigma$ for the pebble flux and expressed $\epsilon$ in terms of the dimensionless accretion rates of \Tb{PA-rates} and the radial drift expression \eqp{vdrift}. Hence, $\epsilon$ can be expressed in terms of dimensionless quantities, $\epsilon=\epsilon(q,\tau_s,h_\mathrm{peb},\dots)$. \Tb{PA-rates} also lists $\epsilon$ in the 2D and 3D limits.  In addition, we can define a mass
\begin{equation}
    M_\mathrm{needed} = \frac{m}{\epsilon}
    = t_\mathrm{growth} \dot{M}_\mathrm{flux}
    \label{eq:needed}
\end{equation}
as the typical mass in pebbles required to grow a planet beyond mass $m$. In \fg{epsmosaic} this quantity is plotted in units of Earth masses for a solar-mass star and $t_\mathrm{growth}$ assumes a mass flux of $100\,m_\oplus\,\mathrm{yr}^{-1}$. In calculating $\epsilon$ we have used expressions and fits applicable for both small and large pebbles \citep{LiuOrmel2018,OrmelLiu2018,HuangOrmel2023} as well as the expressions above for the isolation mass. The pebble scaleheight is calculated according to \citet{DubrulleEtal1995} 
\begin{equation}
    h_\mathrm{peb} = \sqrt{\frac{\delta_z}{\delta_z +\tau_s}} h_\mathrm{gas}
    \label{eq:hpeb}
\end{equation}
where $\delta_z$ represents the turbulent diffusivity normalized by $c_s^2/\Omega_K$.

In \fg{epsmosaic}a we can identify the accretion regimes outlined in \fg{regimes}. It shows that growth by ballistic accretion of pebbles in the Safronov limit is inefficient to entirely ineffective. However, growth accelerates dramatically for masses beyond the $m_\mathrm{init}$ threshold. The chosen parameters, corresponding approximately to conditions representative at 5 au, show that growing a Lunar-mass embryo into an Earth-mass planet requires up to $100\,m_\oplus$ in pebbles of aerodyanmical size in the range $10^{-2} \lesssim \tau_s \lesssim 0.1$.  Thus, while pebble accretion can be fast, it is far from efficient. \Fg{epsmosaic}a also shows that smaller $\tau_s$ pebbles are more efficient building blocks than their larger $\tau_s\sim1$ counterparts, as the reduced drift of smaller pebbles make them more likely to encounter the planet. Very small pebbles, however,  enter the 3D regime, where $\epsilon$ is lower (see \Tb{PA-rates}). Peak efficiencies are obtained at the transitions between the 2D and 3D accretion limit, indicated in \fg{epsmosaic} by the red dashed lines. High efficiencies can also be obtained with large pebbles ($\tau_s>1$), which settle into increasingly thinner layers with reduced radial drift. If present in sufficiently large numbers, these large pebbles become excellent building blocks for planetary growth.

The other panels in \fg{epsmosaic} illustrate how $M_\mathrm{needed}$ depends on the disk properties encapsulated by the parameters $\delta_z$, $\alpha_R$, $h_\mathrm{gas}$ and $\eta$. In \fg{epsmosaic}b it can be seen that a more turbulent disk increases $M_\mathrm{needed}$ considerably for small $\tau_s$, reflecting a shift towards 3D accretion. In the inner disk (\fg{epsmosaic}c) accretion efficiencies are much higher, even for small $\tau_s$, because $h_\mathrm{gas}$ and $\eta$ are lower.  
\textit{In situ} formation of super-Earths/sub-Neptunes is an option. 
The higher accretion rates for large pebbles ($\tau_s\gtrsim10$) in \fg{epsmosaic}c is due to the ballistic accretion channel. In the inner disk, the ratio between the planet's physical and Hill radius increases (higher $\alpha_R$), which increases the probability of a hit.

In the outer disk (\fg{epsmosaic}d) pebble accretion becomes increasingly inefficient. Even though with $\delta_z=10^{-4}$ the disk is rather quiescent, growing planets beyond Earth masses requires ${\gg}10^2\,m_\oplus$ in pebbles. Another problem is that the pebble accretion onset mass $R_\mathrm{init}$ \eqp{Rinit} amounts to bodies of size similar to Pluto, if not Mars. This suppression of pebble accretion arises due to the higher $\eta$ (or gas scaleheight) which affects both the initiation mass ($R_\mathrm{init}\propto\eta^3 \propto h_\mathrm{gas}^6$) and the efficiency ($\epsilon \sim \eta^{-1}$). Although sensitive to the disk model---with colder disk providing more favorable conditions---the efficacy of pebble accretion is doubtful. To render pebble accretion efficient in the (far) outer disk the pebbles' drift velocity must be reduced, which can happen in rings (see below). 

\subsection{Comparison to Planetesimal Accretion}
\begin{table}[t]
    \centering
    \caption{\label{tab:PL-rates}Asymptotic expressions for planetesimal accretion rates.}
    \begin{tabular}{l@{\qquad}r@{}l@{\qquad}r@{}l@{\qquad}l}
    \hline\noalign{\smallskip}
    &   \multicolumn{4}{c}{$\displaystyle \frac{\dot{m}}{\Sigma r^2 \Omega_K}$}          &   validity             \\
    &   \multicolumn{2}{l}{2D}                            & \multicolumn{2}{c}{3D}                   \\
        \hline \noalign{\smallskip}
        geometrical         &   $1.1$&$\alpha_R eq^{1/3}$         & $0.37$          & $\displaystyle \frac{\alpha_R^2 eq^{2/3}}{i} $        & $e\gtrsim q^{1/3} \alpha_R^{-1/2}$ \\[1em]    
        dispersion          &   $2.4$&$\alpha_R^{1/2} q^{2/3} $   & $(1.5{-}1.9)$   & $\displaystyle \frac{\alpha_R q^{4/3} }{ie}$          & $q^{1/3} \lesssim e \lesssim q^{1/3} \alpha_R^{-1/2}$ \\[1em]
        --- $i=e/2$         &       &                           & $6.2$           & $\displaystyle \frac{\alpha_R q^{2/3} }{e_H^2} $          & $e_H \gtrsim 1$   \\[1em]
        shear               &   $5.3$&$\alpha_R^{1/2} q^{2/3} $   & $6.9$           & $\displaystyle \frac{\alpha_R q}{i}$          &   $e \lesssim q^{1/3}$ \\ [0.4em]
\noalign{\smallskip}\hline\noalign{\smallskip}
\end{tabular}
\small
\flushleft
\textbf{Notes.} Accretion rates are given in units of $\Sigma r^2 \Omega_K$ in terms of: $e$ (planetesimal eccentricity), $i$ (planetesimal inclination), $q = m/m_\star$, $\alpha_R=R/R_\mathrm{Hill}$. Expressions for the geometrical and dispersion-dominated (d.d.) limit follow \citet{GreenzweigLissauer1990} and are \textit{not} averaged over a velocity distribution. In the 3D limit the numerical prefactor decreases from 1.9 at $i \ll e$ to $1.5$ at $i/e=0.5$.  For $i/e=0.5$ we also provide the 3D, dispersion-dominated expression in terms of Hill eccentricity $e_H = (3/q)^{1/3}e$, which is by definition ${>}1$ in this limit. The shear-dominated (s.d.) expressions follow \citet{IdaNakazawa1989}.
\end{table}

A frequently-discussed topic is how pebble accretion fares against planetesimal accretion. Unfortunately, this question depends strongly on the amount of accreting material present ($\Sigma$---now split over planetesimals or pebbles) and addressing the question which accretion mechanism dominates greatly depends on the disk model. It is therefore largely a matter of comparing apples to oranges. Still I will make a few comments.

In \Tb{PL-rates} I have compiled the specific \textit{planetesimal} accretion rates in the same way as the accretion rates for pebble accretion (\Tb{PA-rates}). One may see similarities between the expressions with $\alpha_R=R/R_H$ taking the role of $\tau_s$ and the planetesimal's inclination $i$ taking the role of the pebble aspect ratio. 
For planetesimal-driven growth it is the 3D, dispersion-dominated limit that usually applies, with the collisional cross section enhanced by the Safronov factor $\Theta = (v_\mathrm{esc}/\Delta v)^2$. A key characteristics of planetesimal accretion is that the planetesimal eccentricity grows along with the planet's mass (viscous stirring), causing growth to slow down with increasing mass, i.e., the $4/3$ power dependence of $\dot{m}$ on $q$ is misleading as $e$ and $i$ increase concommitently. Eccentricities can become even higher when gravitational stirring by turbulence-induced density inhomogeneities becomes important \citep{IdaEtal2008,OrmelOkuzumi2013}.

Specifically, the pebble and planetesimal growth timescales, assuming the dispersion-dominated (for planetesimals) and 3D limit (for pebbles) are:
\begin{align}
    \label{eq:tgrowthPA}
    t_\mathrm{grw-pebbles}^{3D} &= 0.21 \frac{m_\star}{\Sigma_\mathrm{peb} r^2} \frac{h_\mathrm{peb}}{\tau_s} \Omega_K^{-1} \\
    \label{eq:tgrowthPL}
    t_\mathrm{grw-pltsm}^\mathrm{d.d.} &= 0.16 \frac{m_\star}{\Sigma_\mathrm{pltsm} r^2} \frac{q^{1/3}e_H^2 }{\alpha_R} \Omega_K^{-1}
\end{align}
where $e_H \equiv ev_K/v_H = e (3/q)^{1/3}$ is the Hill eccentricity, which could range from ${\approx}5$ in the inner disk to ${\approx}10$ in the outer disk, dependent on gas density and planetesimal size \citep{KokuboIda2002}.   For pebble accretion, turbulent aerodynamical diffusion increases $t_\mathrm{grw}$ through increasing $h_\mathrm{peb}$ up to the point where it approaches the gas scaleheight. 

In the inner disk, both \eqs{tgrowthPA}{tgrowthPL} generally yield values shorter than the disk's lifetime. As a result, the excitation state of planetesimals will not matter much. Even in scenarios where gravitational focusing is negligible (e.g., $\Theta=1$, corresponding to $e_H^2 \sim \alpha_R^{-1}$), the growth timescale 
for an Earth-mass planet at 1\,au evaluates to ${\sim}1\,\mathrm{Myr}$, assuming $\Sigma r^2/m_\star = 10^{-5}$. 
Therefore, in the inner disk embryos are likely to accrete all planetesimals---achieving isolation---before the gas disk disperses.  
For pebble accretion, on the other hand, the key issue is that their inefficiency: far from all pebbles will be accreted. Aerodynamical sizes are likely to be small due to reduced fragmentation threshold (but see \citealt{MusiolikWurm2019}). \Fg{epsmosaic}c illustrates that a sustained influx of pebbles is necessary. Since the pebbles mass reservoir typically resides in the outer disk, any obstacle---e.g., pressure bump, planetesimal belt \citep{GuillotEtal2014}, or pebble-accreting planets---upstream would prevent these pebbles from reaching the inner disk.

In the outer disk \eqs{tgrowthPA}{tgrowthPL} show that pebble accretion is more powerful than planetesimal accretion, for the following reasons: (i) $\tau_s$ usually is not too small; (ii) $\alpha_R$ is low, unless atmospheric capture of planetesimals is important; and (iii) Hill eccentricities of planetesimals are higher due to reduced gas damping. In addition, planetesimals are more easily ejected from the system. The main obstacles for pebble accretion are again the rather low efficiencies and the initiation mass threshold \eqp{Rinit}. It is a matter of some debate whether a planetesimal formation mechanism such as the streaming instability can produce sufficiently massive bodies to meet the $R_\mathrm{init}$ threshold---at 40\,au this would amount to Pluto-mass. Perhaps, after planetesimal \textit{formation}, growth was first driven by planetesimal accretion, until the point where planets have become big enough to accrete pebbles \citep{LevisonEtal2015,LiuEtal2019i,AlibertEtal2018,SchoonenbergEtal2019,BitschIzidoro2023}.%In these and other models growth is first driven by planetesimal accretion .

One environment where pebble accretion clearly stands out are particle-loaden rings, as seen with ALMA. The concentration of pebbles inside these rings could be due to pressure maxima ($\eta\approx0$) or due to particle pileups ($Z_\mathrm{mid}\gg1$). In both cases pebbles travel on Keplerian orbits and $\Delta v$ is much reduced, such that the $R_\mathrm{init}$ thresholds becomes a lesser issue. Accretion proceeds in the 2D limit at rates that will be extremely rapid:
\begin{equation}
%   t_\mathrm{grw-pebbles}^\mathrm{2d,shear} = \frac{m^{1/3} r}{M_\mathrm{ring}\delta_w \tau_s^{2/3}}   \rightarrow \dots
   t_\mathrm{grw-pebbles}^\mathrm{2d,shear} = 
   0.059\,\mathrm{Myr} \ 
   \frac{\delta_w}{\tau_s^{2/3}}
   \left( \frac{m}{m_\oplus} \right)^{1/3} 
   \left( \frac{M_\mathrm{ring}}{10\,m_\oplus} \right)^{-1}
  %\left( \frac{\delta_w}{0.1} \right)
  %\left( \frac{\tau_s}{10^{-2}} \right)^{-2/3} 
   \left( \frac{m_\star}{m_\odot} \right)^{-5/6} 
   \left( \frac{r}{50\,\mathrm{au}} \right)^{3/2}
  %t_\mathrm{grw-pebbles}^\mathrm{2d,shear} = 0.13\,\mathrm{Myr}\
  %(m)_{\oplus}^{1/3} (M_\mathrm{ring})_\mathrm{10\,m\oplus}^{-1} (\delta_w)_{0.1}^{-1} (\tau_s)^{-2/3}_{0.01}
  %(m_\star)^{1/6}_\odot r_{50\,\mathrm{au}}^{1/2}
    \label{eq:tgrw-2d-shear}
\end{equation}
where we used \eq{2D-shear} and assumed the ring contains a mass $M_\mathrm{ring}$ in pebbles spread out over an annulus of width $\delta_w r$. This shows planets can form, by pebble accretion, at the distances where direct imaging planets are observed. While planetesimal-loaden rings in the inner disk would be able to spawn planets, only pebble accretion is a viable mechanism to do so in the outer disk \citep{LeeEtal2022,LauEtal2022,JiangOrmel2023}. 

\subsubsection{Application}
Pebble accretion has been applied in a variety of settings. In the solar system, it is invoked to explain the emergence and architecture of the gas and ice giants \citep{LevisonEtal2015}, the masses and compositions of the inner solar system planets \citep{LevisonEtal2015i,JohansenEtal2021}, as well as the properties of the Jovian satellite system \citep{ShibaikeEtal2019,RonnetJohansen2020,MadeiraEtal2021}.  Pebble accretion has been proposed to explaining specific exoplanet systems, such as TRAPPIST-1 \citep{SchoonenbergEtal2019}, and circumbinary planets such as Kepler-16 \citep{ColemanEtal2023}. Recently, it has been suggested to be responsible for certain features in protoplanetary disks, such as the compositional makeup of molecular rings \citep{JiangEtal2023} and the sharpness of continuum rings \citep{HuangEtal2024i}. Finally, pebble accretion is invoked to explain the statistical properties of exoplanets using planet population synthesis approaches \citep{BitschEtal2019i,LambrechtsEtal2019,LiuEtal2019ii,VenturiniEtal2024}.

A complete review of these and other works is beyond the scope of this chapter. I refer the interested reader to recent review articles by, for instance, \citet{LiuJi2020} and \citet{DrazkowskaEtal2023}.

\section{Conclusion}

With this concise review I aim to have clarified the following:

\begin{itemize}
    \item Pebbles are defined in terms of the dimensionless stopping time $\tau_s$, its aerodynamical size. Centered around $\tau_s=1$, they are particles that have the ability to drift over significant distances in disks.
    \item There is observational evidence that a large fraction of the solid mass budget in disks once existed in the form of pebble-sized particles.
    \item Pebble accretion requires a dissipative medium (gas). In the settling mechanism, particles are captured within the Hill sphere of the planet, from which they cannot escape. This contrast ballistic accretion, which relies on hitting a surface. 
    \item The settling mechanism starts to dominate over ballistic accretion at a mass $m_\mathrm{init}$ (the mass corresponding to \eq{Rinit}) and becomes fully operational at $m_\ast$ \eqp{mast}. The settling condition will also start to fail for $\tau_s>1$. There is an abrupt change in the accretion rate at $m_\mathrm{init}$.
    \item Pebble accretion cross sections can be large, but in smooth disks growth is not necessarily efficient due to the strong drift of pebbles. A useful quantity to describe pebble accretion is the accretion probability $\epsilon$.  
    \item The pebble accretion rate is determined by the approach velocity $\Delta v$, the planet mass, $\tau_s$, and the pebble scaleheight. These properties determine a variety of accretion regimes (2D vs 3D; shear vs headwind), summarized in \fg{regimes} and \Tb{PA-rates}.
    \item Accretion of aerodynamically large pebbles ($\tau_s>1$, but still drifting significantly) can become very effective as these particles settle to the midplane and approach at low velocity. Early planet atmospheres further boost the accretion probability. 
    \item In realistic situations, pebble accretion is likely to operate alongside planetesimal accretion. In the outer disk pebble accretion, once it is operational, is arguably more effective. 
    \item Dense particulate rings, arising from local pressure maxima or particle pileups---likely the conditions of continuum rings seen with ALMA---can spawn planets through pebble accretion at large distances (several tens of au).
\end{itemize}

\runinhead{Acknowledgments}
I acknowledge support by the National Natural Science Foundation of China (grants no. 12250610189 and 12233004). I appreciate helpful comments by Shuo Huang, Haochang Jiang and Nicolas Kaufmann.

\bibliographystyle{spbasicHBexo}
\bibliography{ads,arXiv}

\end{document}